\magnification \magstep1
\raggedbottom
\openup 2\jot
\voffset6truemm
\centerline {\bf QUALITATIVE PROPERTIES OF THE DIRAC EQUATION}
\centerline {\bf IN A CENTRAL POTENTIAL}
\vskip 1cm
\leftline {Giampiero Esposito$^{1,2}$ and Pietro Santorelli$^{2,1}$}
\vskip 0.3cm
\noindent
{\it ${ }^{1}$Istituto Nazionale di Fisica Nucleare, Sezione
di Napoli, Mostra d'Oltremare Padiglione 20, 80125 Napoli, Italy}
\vskip 0.3cm
\noindent
{\it ${ }^{2}$Universit\`a di Napoli Federico II, Dipartimento
di Scienze Fisiche, Complesso Universitario di Monte S. Angelo,
Via Cintia, Edificio G, 80126 Napoli, Italy}
\vskip 1cm
\noindent
{\bf Abstract.}
The Dirac equation for a massive spin-${1\over 2}$ field in a
central potential $V$ in three dimensions is studied without
fixing a priori the functional form of $V$. The second-order 
equations for the radial parts of the spinor wave function are
shown to involve a squared Dirac operator for the
free case, whose essential 
self-adjointness is proved by using the Weyl limit point-limit circle
criterion, and a `perturbation' 
resulting from the potential. One then finds that
a potential of Coulomb type in the Dirac equation leads to a
potential term in the above second-order equations which is not
even infinitesimally form-bounded with respect to the free 
operator. Moreover, the conditions ensuring essential
self-adjointness of the second-order operators in the interacting
case are changed with respect to the free case, i.e. they are
expressed by a majorization involving the parameter in the Coulomb
potential and the angular momentum quantum number. The same methods
are applied to the analysis of coupled eigenvalue equations when
the anomalous magnetic moment of the electron is not neglected.
\vskip 100cm
\leftline {\bf 1. Introduction}
\vskip 0.3cm
\noindent
In the same year when Dirac derived the relativistic wave equation
for the electron [1], the work of Darwin and Gordon had already
exactly solved such equation in a Coulomb potential in three spatial
dimensions [2, 3]. Since those early days, several efforts have
been produced in the literature to solve the Dirac equation with
other forms of central potentials, until the recent theoretical 
attempts to describe quark confinement [4--8]. In the present paper we
study the mathematical foundations of the eigenvalue problem for a
massive spin-${1\over 2}$ field in a central potential $V(r)$ on
${\bf R}^{3}$, without specifying a priori which function we choose
for $V(r)$. In other words,
we prefer to draw conclusions on $V(r)$ from a careful
mathematical investigation.

By doing so, we hope to elucidate the general framework of 
relativistic eigenvalue problems on the one hand, and to develop
powerful tools to understand some key features of central potentials
on the other. For this purpose, in section 2 we focus on the 
radial parts of the spinor wave function, casting the corresponding
second-order differential operators 
in a convenient form for the subsequent analysis.
In section 3, the Weyl limit point-limit circle criterion [9] is
used to prove that the squared Dirac operator for the free problem
is essentially self-adjoint on the set $C_{0}^{\infty}(0,\infty)$ of
smooth functions on $(0,\infty)$ with compact support away from
the origin. In section 4 some boundedness 
criteria for perturbations [9, 10]
are first described and then applied when the potential in the
original Dirac equation consists of terms of Coulomb and/or linear
type. The effects of the anomalous magnetic moment of the electron
are studied in section 5.
Concluding remarks and open problems are presented in section 6. 
\vskip 0.3cm
\leftline {\bf 2. Second-order equations for stationary states}
\vskip 0.3cm
\noindent
For a charged particle with spin in a central field, the 
angular momentum operator and the parity operator with respect
to the origin of the coordinate system commute with the
Hamiltonian. Thus, states with definite energy, angular
momentum and parity occur. The corresponding spinor
wave function for stationary states reads [11, 12]
$$
\psi=\pmatrix{\varphi \cr \chi \cr}
=\pmatrix{g(r) \Omega_{j,l,m} \cr 
(-1)^{{1+l-l' \over 2}}f(r) \Omega_{j,l',m} \cr}
$$
where $\Omega_{j,l,m}$ and
$\Omega_{j,l',m}$ are the spinor harmonics defined, for example,
in [11, 12], and $l=j \pm {1\over 2}, l'=2j-l$.

The stationary Dirac equation in a central potential $V(r)$ takes
the form ($m_{0}$ being the rest mass of the particle of linear
momentum $\vec p$)
$$
\pmatrix{m_{0}c^{2}+V(r) &  {\vec \sigma} \cdot {\vec p} \cr
{\vec \sigma} \cdot {\vec p} & -m_{0}c^{2}+V(r) \cr}
\pmatrix{\varphi \cr \chi \cr}
=E \pmatrix{\varphi \cr \chi \cr}
$$
and leads eventually to the following coupled system of first-order
differential equations (having defined $F(r) \equiv rf(r)$ and
$G(r) \equiv rg(r)$):
$$
\left({d\over dr}+{k\over r} \right)G(r)=(\lambda_{1}-W(r))F(r)
\eqno (2.1)
$$
$$
\left(-{d\over dr}+{k\over r}\right)F(r)
=(-\lambda_{2}-W(r))G(r)
\eqno (2.2)
$$
where $k=-l-1$ (if $j=l+{1\over 2}$) or 
$l$ (if $j=l-{1\over 2}$), and we have defined
$$
W(r) \equiv {V(r)\over {\hbar}c}
\eqno (2.3)
$$
$$
\lambda_{1} \equiv {E+m_{0}c^{2}\over {\hbar}c}
\eqno (2.4)
$$
$$
\lambda_{2} \equiv {-E+m_{0}c^{2}\over {\hbar}c}.
\eqno (2.5)
$$
Equation (2.1) yields a formula for $F(r)$ which, upon insertion
into Eq. (2.2), leads to the second-order equation
$$
\left[{d^{2}\over dr^{2}}+p(r){d\over dr}+q(r)\right]G(r)=0
\eqno (2.6)
$$
where
$$
p(r) \equiv {W'(r)\over (\lambda_{1}-W(r))}
\eqno (2.7)
$$
$$
q(r) \equiv -{k(k+1)\over r^{2}}+{k\over r}p(r)+W^{2}(r)
+(\lambda_{2}-\lambda_{1})W(r)-\lambda_{1}\lambda_{2}.
\eqno (2.8)
$$
Equation (2.6) should be supplemented by the boundary condition
$G(0)=0$. It then describes a Sturm--Liouville equation non-linear
in the spectral parameter. In [13], the equivalence has been proved
of the radial Dirac equations (2.1) and (2.2) to the 
parameter-dependent Sturm--Liouville equation (2.6) (the parameter
$\lambda$ used in [13] corresponds to our $E$, and $m_{0}c^{2}=1$
units are used therein). By equivalence we mean that, under suitable
assumptions on the potential, the function $G$ solving Eq. (2.6) is
found to belong to the prescribed space $H_{0}^{1}({\bf R}_{+})$,
i.e. the space of absolutely continuous functions on $[0,\infty)$
which are square-integrable on ${\bf R}_{+}$ jointly with their
first derivative and vanish at the origin.
Now we can use a well known technique to transform Eq. (2.6) into
a second-order equation where the coefficient of ${d\over dr}$
vanishes. This is achieved by defining the new function $\Omega$ 
such that [14]
$$
\Omega(r) \equiv G(r) \exp {1\over 2} \int p(r)dr.
\eqno (2.9)
$$
In the few cases where exact analytic formulae are available in
the literature one studies indeed Eq. (2.6) and its countepart
for $F$ (see Eq. (2.13)). However, Eq. (2.9) has the advantage of
leading to a second-order equation for $\Omega$ in a form as close
as possible to `perturbations' of Schr\"{o}dinger operators, and
is hence preferred in our paper devoted to qualitative and
structural properties. In non-relativistic quantum mechanics, such
a method leads to a unitary map [9] transforming the radial
Schr\"{o}dinger equation in a central potential into an equation 
involving a radial Schr\"{o}dinger operator
$-{d^{2}\over dr^{2}}+U(r)$ acting on square-integrable functions
on ${\bf R}_{+}$ which vanish at the origin. In our relativistic
eigenvalue problem the transformation of the Hilbert space of
square-integrable functions is no longer unitary, but remains
of practical value. All non-linear properties of the resulting
Sturm--Liouville boundary-value problem are in fact encoded into
a single function playing the role of parameter-dependent
potential term (see below), rather than two functions $p$ and $q$
as in (2.6)--(2.8). The function
$\Omega$ is then found to obey the differential equation
$$
\left[-{d^{2}\over dr^{2}}+{l(l+1)\over r^{2}}
+P_{W,E}(r)\right]\Omega(r)={(E^{2}-m_{0}^{2}c^{4})\over
{\hbar}^{2}c^{2}}\Omega(r)
\eqno (2.10)
$$
having defined 
$$ \eqalignno{
P_{W,E}(r) & \equiv -W^{2}(r)+{1\over 2}{W''\over (\lambda_{1}-W)}
+{3\over 4}{\left({W' \over \lambda_{1}-W}\right)}^{2} \cr
& -{k\over r}{W'\over (\lambda_{1}-W)}
+{2E \over {\hbar}c}W(r).
&(2.11)\cr}
$$
Such an equation may be viewed as follows: since the potential $W$
`perturbs' the `free' problem for which $W$ vanishes in Eqs.
(2.1) and (2.2), in the corresponding second-order equation (2.10)
one deals with a `free operator'
$$
A_{r}^{l} \equiv -{d^{2}\over dr^{2}}+{l(l+1)\over r^{2}}
\; \; {\rm for} \; {\rm all} \; l =0,1,...
\eqno (2.12)
$$
perturbed by the multiplication operator $P_{W,E}(r)$ defined in
(2.11). An interesting programme is therefore emerging
at this stage:
\vskip 0.3cm
\noindent
(i) First, prove (essential) self-adjointness of the `free' 
operator $A_{r}^{l}$ on a certain domain.
\vskip 0.3cm
\noindent
(ii) Second, try to understand whether the operator
$A_{r}^{l}+P_{W,E}(r)$
in Eq. (2.10) remains self-adjoint on the same domain. If this
condition is too restrictive, try to derive all properties 
of this `perturbed' second-order operator.

If one first uses Eq. (2.2) to relate $G(r)$ to ${dF\over dr}$
and $F$, one finds instead the Sturm--Liouville equation
(cf [13])
$$
\left[{d^{2}\over dr^{2}}+{\widetilde p}(r){d\over dr}
+{\widetilde q}(r)\right]F(r)=0
\eqno (2.13)
$$
supplemented by the boundary condition $F(0)=0$,
where (cf (2.7) and (2.8))
$$
{\widetilde p}(r) \equiv -{W'(r)\over (\lambda_{2}+W(r))}
\eqno (2.14)
$$
$$
{\widetilde q}(r) \equiv -{k(k-1)\over r^{2}}
-{k\over r}{\widetilde p}(r)+W^{2}(r)
+(\lambda_{2}-\lambda_{1})W(r)-\lambda_{1}\lambda_{2}.
\eqno (2.15)
$$
Thus, after defining (cf (2.9))
$$
{\widetilde \Omega}(r) \equiv F(r) \exp {1\over 2}
\int {\widetilde p}(r)dr
\eqno (2.16)
$$
one finds for ${\widetilde \Omega}(r)$ the second-order
differential equation
$$
\left[-{d^{2}\over dr^{2}}+{k(k-1)\over r^{2}}
+{\widetilde P}_{W,E}(r)\right]{\widetilde \Omega}(r)
={(E^{2}-m_{0}^{2}c^{4})\over {\hbar}^{2}c^{2}}
{\widetilde \Omega}(r)
\eqno (2.17)
$$
having now defined (cf (2.11))
$$ \eqalignno{
{\widetilde P}_{W,E}(r) & \equiv -W^{2}(r)-{1\over 2}
{W'' \over (\lambda_{2}+W)}+{3\over 4}
{\left({W' \over \lambda_{2}+W}\right)}^{2} \cr
&-{k\over r}{W' \over (\lambda_{2}+W)}
+{2E \over {\hbar}c}W(r) .
&(2.18)\cr}
$$
Since $k=-l-1$ if $j=l+{1\over 2}$, and $k=l$ if $j=l-{1\over 2}$,
the `free' operator in Eq. (2.17) reads now
$$
{\widetilde A}_{r}^{l} \equiv
-{d^{2}\over dr^{2}}+{(l+1)(l+2)\over r^{2}}
\; \; {\rm for} \; {\rm all} \; l=0,1,... 
\eqno (2.19a)
$$
$$
{\widetilde A}_{r}^{l} \equiv
-{d^{2}\over dr^{2}}+{l(l-1)\over r^{2}}
\; \; {\rm for} \; {\rm all} \; l=1,2,... \; . 
\eqno (2.19b)
$$
Note that $P_{W,E}(r)$ has a second-order pole at
$\lambda_{1}=W$ (see (2.11)) and ${\widetilde P}_{W,E}(r)$ has
a second-order pole at $\lambda_{2}=-W$. Thus, the analysis of
the interacting case (i.e. with $W(r) \not = 0$) is performed in
section 4 at fixed values of $E$ and away from such singular 
points.
\vskip 0.3cm
\leftline {\bf 3. Weyl criterion for the squared Dirac operator in 
the free case}
\vskip 0.3cm
\noindent
The self-adjointness properties of the free operator (2.12) should
be studied by considering separately the case $l>0$ and the case $l=0$.
For positive values of the quantum number $l$, $A_{r}^{l}$ turns 
out to be essentially self-adjoint. This means, by definition,
that its closure (i.e. the smallest closed extension) is
self-adjoint, which implies that a unique self-adjoint extension of 
$A_{r}^{l}$ exists [15]. 
In general, if several self-adjoint extensions exist, one has to 
understand which one should be chosen, 
since they are distinguished by the physics of the system being
described [9, 15]. This is why it is so desirable to make sure that
the operator under investigation is essentially self-adjoint.
We here rely on a criterion due to
Weyl, and the key steps are as follows [9]. 

The function $V$ is in the {\it limit circle} case at zero if for some,
and therefore all $\lambda$, {\it all} solutions of the equation
$$
\left[-{d^{2}\over dx^{2}}+V(x) \right]\varphi(x)=\lambda \varphi(x)
\eqno (3.1)
$$
are square integrable at zero, i.e. for them
$$
\int_{0}^{a}|\varphi(x)|^{2}dx < \infty
\eqno (3.2)
$$
with finite values of $a$, e.g. $a \in ]0,1]$. If $V(x)$ is not in
the limit circle case at zero, it is said to be in the
{\it limit point} case at zero. The Weyl limit point-limit circle
criterion states that, if $V$ is a continuous real-valued function
on $(0,\infty)$, then the operator
$$
{\cal O} \equiv -{d^{2}\over dx^{2}}+V(x)
\eqno (3.3)
$$
is essentially self-adjoint on $C_{0}^{\infty}(0,\infty)$ if and only
if $V(x)$ is in the limit point case at both zero and infinity. The 
property of being in the limit point at zero relies on [9]
\vskip 0.3cm
\noindent
{\bf Theorem 3.1} Let $V$ be continuous and positive near zero. If
$$
V(x) \geq {3\over 4}x^{-2}
\eqno (3.4)
$$
near zero, then $\cal O$ is in the limit point case at zero.

The limit point property at $\infty$ means that the limit
circle condition at $\infty$ is not fulfilled, i.e. the condition
$$
\int_{a}^{\infty}|\varphi(x)|^{2}dx < \infty
\eqno (3.5)
$$
does not hold. To understand when this happens, one can use [9]
\vskip 0.3cm
\noindent
{\bf Theorem 3.2} If $V$ is differentiable on $(0,\infty)$ and
bounded above by a parameter $K$ on $[1,\infty)$, and if
$$
\int_{1}^{\infty}{dx\over \sqrt{K-V(x)}}=\infty
\eqno (3.6)
$$
$$
V'(x)|V(x)|^{-{3\over 2}} \; \; {\rm is} \; \; 
{\rm bounded} \; \; {\rm near} \; \; \infty
\eqno (3.7)
$$
then $V(x)$ is in the limit point case at $\infty$.
\vskip 0.3cm
\noindent
Thus, a necessary and sufficient condition for the existence of a
unique self-adjoint extension of $\cal O$ is that its eigenfunctions
should fail to be square integrable at zero and at $\infty$. 
Powerful operational criteria are provided by the check of (3.4),
(3.6) and (3.7), which only involve the potential.

In our problem, for all $l \geq 1$, the `potential'
${\widetilde V}_{l}(r) \equiv {l(l+1)\over r^{2}}$ is of course
in the limit point at zero, since the inequality (3.4) is then
satisfied. Moreover, ${\widetilde V}_{l}(r)$ is differentiable on
$(0,\infty)$, bounded above by $\chi_{l} \equiv l(l+1)$ on
$[1,\infty)$, and such that
$$
\int_{1}^{\infty}{dx\over \sqrt{\chi_{l}-{\widetilde V}_{l}(x)}}
={1\over \sqrt{l(l+1)}} \int_{1}^{\infty}{x\over \sqrt{x^{2}-1}}dx
=\infty
\eqno (3.8)
$$
$$
{\widetilde V}_{l}'(r) |{\widetilde V}_{l}(r)|^{-{3\over 2}}
=-{2\over \sqrt{l(l+1)}} \; \; {\rm for} \; {\rm all} \; r.
\eqno (3.9)
$$
Hence all conditions of theorem 3.2 are satisfied, which implies that
${\widetilde V}_{l}(r)$ is in the limit point at $\infty$ as well. 
By virtue of the Weyl limit point-limit circle criterion, the free
operator $A_{r}^{l}$ defined in (2.12) is then essentially 
self-adjoint on $C_{0}^{\infty}(0,\infty)$ for all $l > 0$. 

When $l=0$, however (for which $k=-1$), 
$A_{r}^{l}$ reduces to the operator 
$-{d^{2}\over dr^{2}}$, which has deficiency indices $(1,1)$. Recall
that for an (unbounded) operator $B$ with adjoint $B^{\dagger}$, 
deficiency indices are the dimensions of the spaces of solutions of 
the equations $B^{\dagger}u=\pm iu$. More precisely, one defines 
first the deficiency sub-spaces ($D(B^{\dagger})$ being the domain
of $B^{\dagger}$)
$$
{\cal H}_{+}(B) \equiv \left \{ u \in D(B^{\dagger}):
B^{\dagger}u=iu \right \}
\eqno (3.10)
$$
$$
{\cal H}_{-}(B) \equiv \left \{ u \in D(B^{\dagger}):
B^{\dagger}u=-iu \right \}
\eqno (3.11)
$$
with corresponding deficiency indices
$$
n_{+}(B) \equiv {\rm dim} \; {\cal H}_{+}(B)
\eqno (3.12)
$$
$$
n_{-}(B) \equiv {\rm dim} \; {\cal H}_{-}(B).
\eqno (3.13)
$$ 
The operator $B$ is self-adjoint if and only if 
$n_{+}(B)=n_{-}(B)=0$, but has self-adjoint extensions provided that
$n_{+}(B)=n_{-}(B)$ [9, 15]. 
In our case, half of the solutions of the
equations ${(A_{r}^{0})}^{\dagger}u=\pm iu$ are square-integrable on
${\bf R}_{+}$, which implies that $n_{+}(A_{r}^{0})=
n_{-}(A_{r}^{0})=1$. This is easily proved because such equations
with complex eigenvalues reduce to the ordinary differential
equation [9]
$$
-{d^{2}\over dr^{2}}{\rm e}^{\alpha r}=i {\rm e}^{\alpha r}
\eqno (3.14)
$$
and
$$
-{d^{2}\over dr^{2}}{\rm e}^{\omega r}=-i {\rm e}^{\omega r}.
\eqno (3.15)
$$
In the former case, on setting $\alpha=\rho {\rm e}^{i \theta}$,
with $\rho$ and $\theta \in {\bf R}$, one finds
$\rho=\pm 1, \theta=-{\pi \over 4}$, which leads to the two roots
of the equation $-\alpha^{2}=i$:
$$
\alpha_{1}={1\over \sqrt{2}}-{i\over \sqrt{2}}
\eqno (3.16)
$$
$$
\alpha_{2}=-{1\over \sqrt{2}}+{i\over \sqrt{2}}.
\eqno (3.17)
$$
In the latter case, $\omega$ solves the algebraic equation 
$\omega^{2}=i$, and hence one finds the roots
$$
\omega_{1}={1\over \sqrt{2}}+{i\over \sqrt{2}}
\eqno (3.18)
$$
$$
\omega_{2}=-{1\over \sqrt{2}}-{i\over \sqrt{2}}.
\eqno (3.19)
$$
Only the roots $\alpha_{2}$ and $\omega_{2}$ are compatible with
the request of square-integrable solutions of (3.14) and (3.15)
on ${\bf R}_{+}$, and hence one finds $n_{+}(A_{r}^{0})
=n_{-}(A_{r}^{0})=1$ as we anticipated. This property implies that
a one-parameter family of self-adjoint extensions of $A_{r}^{0}$
exists, with domain $D(A_{r}^{0})$ given by
$$ \eqalignno{
\; & D(A_{r}^{0})= \left \{ u \in L^{2}({\bf R}_{+}):
u,u' \in AC_{\rm {loc}}({\bf R}_{+});
u'' \in L^{2}({\bf R}_{+}); \right . \cr
& \left . u(0)=\beta u'(0) \right \}.
&(3.20)\cr}
$$
Here $AC_{\rm {loc}}({\bf R}_{+})$ denotes the set of locally
absolutely continuous functions on the positive half-line, the
prime denotes differentiation with respect to $r$, and $\beta$
is a real-valued parameter. Bearing in mind the limiting form
of Eq. (2.10) when $l=0$ and $W=0$, this means that one is 
studying the case characterized by
$$
\lambda \equiv {(E^{2}-m_{0}^{2}c^{4})
\over {\hbar}^{2}c^{2}} < 0
\eqno (3.21)
$$
for which the square-integrable eigenfunction of
$-{d^{2}\over dr^{2}}$ reads ($\sigma$ being a real constant
to ensure reality of $E$)
$$
u(r)=\sigma \; {\rm e}^{-r \sqrt{| \lambda |}}.
\eqno (3.22)
$$
On defining
$$
(u,v) \equiv \int_{0}^{\infty} u^{*}(r)v(r)dr
$$
the boundary condition in (3.20) is obtained after integrating 
twice by parts in the integral defining the scalar product
$\Bigr(A_{r}^{0}u,v \Bigr)$
to re-express it in the form $\Bigr(u, (A_{r}^{0})^{\dagger}v
\Bigr)$, with $u$ in the domain of $A_{r}^{0}$ and $v$ in the 
domain of the adjoint $(A_{r}^{0})^{\dagger}$. One then finds that
both $u$ and $v$ should obey the boundary condition (3.20). In
the light of (3.20)--(3.22) one obtains the very useful formula
$$
1=-\beta \sqrt{| \lambda |}
\eqno (3.23)
$$
which implies
$$
E^{2}=m_{0}^{2}c^{4}-{{\hbar}^{2}c^{2}\over \beta^{2}}.
\eqno (3.24)
$$
This means that in a relativistic problem a lower limit for 
$\beta^{2}$ (and hence for $|\beta|$) exists, to avoid
having $E^{2} < 0$.

To complete the analysis of squared Dirac operators in the free case,
one has also to consider the operators ${\widetilde A}_{r}^{l}$
defined in (2.19a) and (2.19b). The former has a `potential' term
${(l+1)(l+2)\over r^{2}}$ which is in the limit point case at both 
zero and infinity for all $l \geq 0$. The latter has a 
`potential' term ${l(l-1)\over r^{2}}$ which is in the limit point
at zero with the exception of the value $1$ of the quantum
number $l$, for which ${\widetilde A}_{r}^{l}$ reduces to the
operator $-{d^{2}\over dr^{2}}$, and hence we repeat the logical
steps proving that such an operator has a one-parameter family of
self-adjoint extensions. Once more, 
their domain is given by Eq. (3.20).
\vskip 0.3cm
\leftline {\bf 4. Second-order operators in the interacting case}
\vskip 0.3cm
\noindent
Now we would like to understand whether the general results on
perturbations of self-adjoint operators make it possible to
obtain a better understanding of effects produced by the 
central potential $W(r)$ in Eqs. (2.10) and (2.17) (the essential
self-adjointness of the Dirac Hamiltonian with non-vanishing $W$ is
studied in [16], and several comments can be found in the following
sections). For this purpose, the key steps are as follows [9].
\vskip 0.3cm
\noindent
(i) Let $A$ and $B$ be densely defined linear operators on a
Hilbert space $H$ with domains $D(A)$ and $D(B)$, respectively.
If $D(A) \subset D(B)$ and if, for some $a$ and $b$ in $\bf R$,
$$
\left \| B \varphi \right \| \leq a \left \| A \varphi \right \|
+b \left \| \varphi \right \| \; \; {\rm for} \; {\rm all} \;
\varphi \in D(A)
\eqno (4.1)
$$
then $B$ is said to be {\it $A$-bounded}. The infimum of such $a$ 
is called the {\it relative bound} of $B$ with respect to $A$. If
the relative bound vanishes, the operator $B$ is said to be
{\it infinitesimally small} with respect to $A$.
\vskip 0.3cm
\noindent
(ii) The Kato--Rellich theorem states that if $A$ is self-adjoint,
$B$ is symmetric, and $B$ is $A$-bounded with relative bound
$a < 1$, then $A+B$ is self-adjoint on $D(A)$.
\vskip 0.3cm
\noindent
(iii) If the potential $V$ can be written as
$$
V=V_{1}+V_{2}
\eqno (4.2)
$$
with $V_{1} \in L^{2}({\bf R}^{3})$ and $V_{2} \in 
L^{\infty}({\bf R}^{3})$, and if $V$ is real-valued, then the 
operator $-\bigtriangleup +V(x)$ is essentially self-adjoint on
$C_{0}^{\infty}({\bf R}^{3})$ and self-adjoint on 
$D(-\bigtriangleup)$. As a corollary, the operator 
$-\bigtriangleup -{e^{2}\over r}$ is essentially self-adjoint
on $C_{0}^{\infty}({\bf R}^{3})$.
\vskip 0.3cm
\noindent
(iv) An analogue of the Kato--Rellich theorem exists which can
be used to study the case when $B$ is not $A$-bounded. The result
can be stated after recalling the following definitions.

Let $A$ be a self-adjoint operator on $H$. On passing to a spectral
representation of $A$ with associated measures 
$\left \{ \mu_{n} \right \}_{n=1}^{N}$ on the spectrum of $A$,
so that $A$ is multiplication by $x$ on
the direct sum $\oplus_{n=1}^{N}L^{2}({\bf R},d\mu_{n})$, one
can consider
$$
{\cal I} \equiv \left \{ \left \{ \psi_{n}(x) \right \}_{n=1}^{N}:
\; \sum_{n=1}^{N} \int_{-\infty}^{\infty}|x| 
{|\psi_{n}(x)|}^{2} d\mu_{n} < \infty \right \}
\eqno (4.3)
$$
and hence define, for $\psi$ and $\varphi \in {\cal I}$,
$$
q(\varphi,\psi) \equiv \sum_{n=1}^{N} 
\int_{-\infty}^{\infty}x \varphi_{n}^{*}(x)\psi_{n}(x)
d\mu_{n}.
\eqno (4.4)
$$
Such a $q$ is called the {\it quadratic form} associated with
$A$, and one writes 
$$
Q(A) \equiv {\cal I}.
\eqno (4.5)
$$
The {\it form domain} of the operator $A$ is then, by definition,
$Q(A)$, and can be viewed as the largest domain on which $q$ can
be defined.
\vskip 0.3cm
\noindent
(v) The KLMN theorem states that, if $A$ is a positive self-adjoint
operator and if $\beta(\varphi,\psi)$ is a symmetric quadratic form
on $Q(A)$ such that
$$
|\beta(\varphi,\varphi)| \leq a (\varphi,A \varphi)
+b (\varphi,\varphi) \; \; {\rm for} \; {\rm all} \;
\varphi \in D(A)
\eqno (4.6)
$$
for some $a < 1$ and $b \in {\bf R}$, then there exists a unique
self-adjoint operator $C$ with
$$
Q(C)=Q(A)
\eqno (4.7)
$$
and
$$
(\varphi, C \psi)=(\varphi, A \psi)+\beta(\varphi,\psi)
\; \; {\rm for} \; {\rm all} \; \varphi,\psi \in Q(C).
\eqno (4.8)
$$
Such a $C$ is bounded below by $-b$.
\vskip 0.3cm
\noindent
(vi) If $A$ is a positive self-adjoint operator, and $B$ is a
self-adjoint operator such that
$$
Q(A) \subset Q(B)
\eqno (4.9)
$$
and
$$
|(\varphi,B \varphi)| \leq a (\varphi,A \varphi)
+b(\varphi,\varphi) \; \; {\rm for} \; {\rm all} \;
\varphi \in D(A)
\eqno (4.10)
$$
for some $a>0$ and $b \in {\bf R}$, then $B$ is said to be 
{\it relatively form-bounded} with respect to $A$. Furthermore, if
$a$ can be chosen arbitrarily small, $B$ is said to be 
{\it infinitesimally form-bounded} with respect to $A$.
\vskip 0.3cm
\noindent
(vii) If the operator $B$ is self-adjoint and relatively 
form-bounded, the parameter $a$ being $<1$, with respect to a
positive self-adjoint operator $A$, then the KLMN theorem makes it
possible to define the `sum' $A+B$, although this mathematical
construction may differ from the operator sum. In particular,
$B$ can be form-bounded with respect to $A$ even though the 
intersection of their domains may be the empty set.
\vskip 0.3cm
\noindent
(viii) The KLMN theorem is physically relevant because it leads 
to the definition of Hamiltonians even when the Kato--Rellich 
criterion is not fulfilled. In other words, the request of dealing
with $L^{2}+L^{\infty}$ potentials is too restrictive. For example, 
the potential $V_{\alpha}(r)=-r^{-\alpha}$ belongs to
$L^{2}+L^{\infty}$ only if $\alpha < {3\over 2}$. However, if
$\alpha \in \left[{3\over 2},2 \right)$, one can use the KLMN
theorem because, for all $\alpha < 2$, one can prove that 
$-r^{-\alpha}$ is infinitesimally form-bounded with respect 
to $-\bigtriangleup$ [9]. 
\vskip 0.3cm
In our problem, the `potential' terms in Eqs. (2.10) and (2.17)
are given by (2.11) and (2.18), respectively. If the potential
$W(r)$ is of Coulomb type, i.e. ($\gamma$ being a dimensionful
constant)
$$
W(r)={\gamma \over r}
\eqno (4.11)
$$
the singular behaviour of $P_{W,E}(r)$ as $r \rightarrow 0$
is dominated by (for a {\it fixed value} of $E$)
$$
-{(\gamma^{2}+{1\over 4}+k)\over r^{2}}
$$
and the singular behaviour of ${\widetilde P}_{W,E}(r)$ as
$r \rightarrow 0$ is given instead by (again for a fixed
value of $E$)
$$
-{(\gamma^{2}+{1\over 4}-k)\over r^{2}}.
$$
Thus, as $r \rightarrow 0$, the operators on the left-hand sides
of both (2.10) and (2.17) reduce to
$$
L_{r} \equiv \left[-{d^{2}\over dr^{2}}
+{\left(k^{2}-\gamma^{2}-{1\over 4} \right)\over r^{2}}
\right].
\eqno (4.12)
$$
In the operator $L_{r}$, the coefficient of $r^{-2}$ is no longer
greater than or equal to ${3\over 4}$ (see (3.4)) for the same 
values of $l$ ensuring essential 
self-adjointness of the free problem. The inequality 
$$
k^{2}-\gamma^{2}-{1\over 4} \geq {3\over 4}
\eqno (4.13)
$$
is instead fulfilled by
$$
(l+1)^{2} \geq \gamma^{2}+1 \; \; {\rm for} \; {\rm all} \;
l=0,1,...
\eqno (4.14)
$$
if $k=-l-1$, and by 
$$
l^{2} \geq \gamma^{2}+1 \; \; {\rm for} \; {\rm all} \;
l=1,2,... 
\eqno (4.15)
$$
if $k=l$.
Our result implies that, for all $|k| \geq 2$, essential 
self-adjointness on $C_{0}^{\infty}(0,\infty)$ of the second-order
operators on the left-hand sides of (2.10) and (2.17) 
is obtained provided that $|\gamma| \leq \sqrt{3}$.
This reflects the fact that a Coulomb potential in the first-order
system (2.1) and (2.2) leads to `potential' terms in the 
second-order equations (2.10) and (2.17) which are not even
infinitesimally form-bounded with respect to the squared Dirac
operators in the free case, because both the potential terms 
and the free operators contain terms proportional to $r^{-2}$.
To study the limit point condition at infinity, we try to majorize
the `potential' $P_{W,E}$ obtained from the Coulomb potential 
(4.11), and we find that
$$
|P_{W,E}(r)| \leq 
2{|E \gamma| \over {\hbar}c}
+{\Bigr[|\gamma(1+k)| (\lambda_{1}+|\gamma|)
+{3\over 4}\gamma^{2}\Bigr]\over \lambda_{1}^{2}}
$$
if $r \in [1, \infty)$.
Moreover, the integral (3.6) diverges when $V$ is replaced by
$P_{W,E}$, and the condition (3.7) is fulfilled as well, because
$$
P_{W,E}'(r) |P_{W,E}(r)|^{-{3\over 2}} \; \propto \; 
r^{-{1\over 2}} \; {\rm as} \; r \rightarrow \infty.
$$
The check of (3.6) and (3.7) for ${\widetilde P}_{W,E}$ leads to
the same results, and hence we use the Weyl criterion of section 3
to conclude that, {\it for fixed values} of $E$, essential self-adjointness
on $C_{0}^{\infty}(0,\infty)$ of the second-order operators 
in Eqs. (2.10) and (2.17) holds provided that the inequality
$k^{2}-\gamma^{2} \geq 1$
is satisfied. This rules out $l=0$ in (4.14) and $l=1$ in (4.15).
One then finds that $|\gamma| \leq \sqrt{3}$ as we said before. 

The limiting form (4.12) is not affected by the addition of parts 
linear in $r$ [6, 17, 18] to the right-hand side of (4.11), 
because the singular behaviour of $P_{W,E}(r)$ 
at fixed values of $E$ as
$r \rightarrow 0$ is still dominated by the Coulomb potential.
By contrast, a purely linear potential
$$
W(r)=\Gamma r
\eqno (4.16)
$$
satisfies the request of infinitesimal form-boundedness 
of $P_{W,E}(r)$ with respect to the squared Dirac operators in
the free case, because then the singular behaviour
of $P_{W,E}(r)$ as $r \rightarrow 0$ is expressed by
$-{k\Gamma \over \lambda_{1}} {1\over r}$
and the singular part of ${\widetilde P}_{W,E}(r)$ as
$r \rightarrow 0$ reads
$-{k\Gamma \over \lambda_{2}} {1\over r}$.
However, one might consider linear terms with compact
support, i.e. vanishing for all $r$ greater than some finite
$r_{0}$, or weighted with exponential functions which ensure a
fall-off condition at infinity, e.g. the potential (cf [17])
$$
W(r)={\gamma \over r}+\Gamma r {\rm e}^{-\mu r}
\eqno (4.17)
$$
where $\mu$ is positive. In such a case, the limiting behaviours 
of $P_{W,E}$ as $r \rightarrow 0$
and as $r \rightarrow \infty$ are still dominated by the Coulomb
part in the potential $W$, and hence we find again essential 
self-adjointness on $C_{0}^{\infty}(0,\infty)$ provided that
$k^{2}-\gamma^{2} \geq 1$.

In the physical literature, however, the potential has not been
written in the form (4.17). To achieve quark confinement, a purely
linear term has instead been added to the Coulomb part, considering
also a split of the additional part into Lorentz scalar-type and
Lorentz vector-like potentials. Furthermore, such a vector
contribution is sometimes omitted in a phenomenological analysis,
bearing in mind its non-perturbative nature (since the perturbative
part has instead vector nature) [19]. Needless to say, such 
arguments are not compelling. 
\vskip 0.3cm
\leftline {\bf 5. Inclusion of the anomalous magnetic moment}
\vskip 0.3cm
\noindent
The second-order operators that we have analyzed in the interacting 
case (see again Eqs. (2.10) and (2.17)) are not `squared Dirac
operators' because the eigenvalues of the Dirac operator occur
in their `potential term'. It is therefore important to compare
more carefully the predictions of the second-order equation for
$\Omega$ (and $\widetilde \Omega$) with the results obtained from
squared Dirac operators studied in [16]. The latter are used in
[16] because a theorem ensures that, given the (abstract) Dirac
Hamiltonian
$$
T=\pmatrix{0 & D_{-} \cr D_{+} & 0 \cr}
+ \pmatrix{W_{+} & 0 \cr 0 & W_{-} \cr}
\eqno (5.1)
$$
if one of the operators $D_{-}D_{+}$ or $D_{+}D_{-}$ is essentially
self-adjoint, then the operator $T$ is essentially self-adjoint
as well, where $W_{+}$ and $W_{-}$ take into account the rest mass
and the potential (see theorem 5.9 in [16]).

Let us now consider the effect of the anomalous magnetic moment
$\mu$ of the electron in a central potential $V(r)$. With the
notation of our section 2, the resulting set of coupled eigenvalue
equations is found to be [16]
$$
\left[{d\over dr}+{k\over r}-\mu W'(r)\right]G(r)
=(\lambda_{1}-W(r))F(r)
\eqno (5.2)
$$
$$
\left[-{d\over dr}+{k\over r}-\mu W'(r)\right]F(r)
=(-\lambda_{2}-W(r))G(r)
\eqno (5.3)
$$
which implies, on using again the definition (2.9), that $\Omega(r)$
obeys the second-order equation
$$
\left[-{d^{2}\over dr^{2}}+{l(l+1)\over r^{2}}
+P_{W,E}^{(\mu)}(r)\right]\Omega(r)
={(E^{2}-m_{0}^{2}c^{4})\over {\hbar}^{2}c^{2}}\Omega(r)
\eqno (5.4)
$$
where we have defined (cf Eq. (5.48) in [16])
$$
P_{W,E}^{(\mu)}(r) \equiv P_{W,E}(r)+\mu \left[W''
+{W'}^{2}\Bigr(\mu+(\lambda_{1}-W)^{-1}\Bigr)-2{k\over r}W'
\right].
\eqno (5.5)
$$
For example, if a potential $W$ of Coulomb type is considered, one
finds from (4.11) and (5.5) that the limiting form of the
eigenvalue equation (5.4) as $r \rightarrow 0$ is entirely
dominated by the term proportional to $\mu$. More precisely, in
such a limit Eq. (5.4) reduces to
$$
\left[{d^{2}\over dr^{2}}-{\mu^{2}\gamma^{2}\over r^{4}}
\right]\Omega(r)=0
\eqno (5.6)
$$
which is solved by
$$
\Omega(r)=r \; {\rm e}^{-{\mu \gamma \over r}}.
\eqno (5.7)
$$
An analogous method can be used for 
${\widetilde \Omega}(r)$ defined in (2.16), finding a
parameter-dependent potential 
$$
{\widetilde P}_{W,E}^{(\mu)}(r) \equiv {\widetilde P}_{W,E}(r)
+\mu \left[-W''+{W'}^{2}\left(\mu+{1\over (\lambda_{2}+W)}
\right)-2{k\over r}W' \right]
\eqno (5.8)
$$
which leads again to the limiting behaviour (5.7) when
$W(r)={\gamma \over r}$, but now for ${\widetilde \Omega}(r)$,
as $r \rightarrow 0$. We can therefore see,
in a physically relevant example, that our approach, leading to
second-order equations for $\Omega$ and $\widetilde \Omega$,
recovers qualitative agreement with the analysis in [16],
where it is shown that, no matter how singular is the central 
potential at $r=0$, the Dirac operator is always well defined as
long as $\mu \not = 0$. In other words, 
our formula (5.5) accounts clearly for the
dominating effect of the anomalous magnetic moment with all potentials
diverging at the origin. However, a rigorous result on the relation
between our approach and the squared Dirac operators studied 
in [16] remains an interesting technical problem whenever 
$W(r) \not = 0$ (cf [13]).
\vskip 0.3cm
\leftline {\bf 6. Concluding remarks}
\vskip 0.3cm
\noindent
The contributions of our paper, of technical nature, consist in
the application of analytic techniques that can help to understand
some key qualitative features of
central potentials for the Dirac equation, with emphasis on the
mathematical formulation of relativistic eigenvalue problems.
Although the methods used in our investigation are well known
in the literature, the overall picture remains, to our knowledge,
original (see comments below). 
In particular, we would like to mention the following
points (at the risk of slight repetitions).
\vskip 0.3cm
\noindent
(i) The forms (2.10) and (2.17) 
of the second-order equations for the radial parts of
the spinor wave function, with $P_{W,E}(r)$ and
${\widetilde P}_{W,E}(r)$ defined in (2.11) and (2.18),
respectively, is very convenient if one wants to understand
whether the potential can
affect the self-adjointness domain of the free problem.
\vskip 0.3cm
\noindent
(ii) The identification of the domains of (essential) self-adjointness
of the operators defined in (2.12), (2.19a) and 
(2.19b) is helpful as a first step
towards the problem with non-vanishing potential $W(r)$, and
clarifies the general framework. 
\vskip 0.3cm
\noindent
(iii) A potential of Coulomb type, although quite desirable from a
physical point of view, leads to some non-trivial features with
respect to the non-relativistic case. We have in fact seen that
$P_{W,E}(r)$ and ${\widetilde P}_{W,E}(r)$ fail to be infinitesimally
form-bounded with respect to the squared Dirac operators 
in the free case, if $W(r)$ contains a 
Coulomb term. Moreover, the limit-point condition at
zero for the potential in the second-order operators in the
interacting case is only fulfilled if the inequalities (4.14)
or (4.15) hold. In other words, the essential self-adjointness
on $C_{0}^{\infty}(0,\infty)$ of the second-order operators with
non-vanishing potential is still obtained, but under more restrictive
conditions expressed by (4.14) and (4.15). This may have non-trivial
physical implications: if essential self-adjointness fails to hold,
we know from section 3 that different self-adjoint 
extensions of the second-order operators exist, characterized by 
the choice of regular boundary condition at $r=0$ (cf (3.20)). 
The lowest values of $l$ (for which (4.13) does not hold),
corresponding to the bound states of greater phenomenological 
interest, might therefore find an appropriate mathematical
description within the framework of self-adjoint extensions of
symmetric operators. It remains to be seen how much freedom is
left, on physical ground, to specify the boundary conditions for the
self-adjoint extension.
\vskip 0.3cm
\noindent
(iv) On considering the effect of the anomalous magnetic moment, 
Eq. (2.10) is replaced by Eq. (5.4), with the potential term
defined in (5.5). For all potentials diverging at the origin, the
effect of the anomalous magnetic moment is then dominating
as $r \rightarrow 0$.
\vskip 0.3cm
Indeed, as far as the Dirac operator is concerned, one can prove
its essential self-adjointness on $C_{0}^{\infty}\Bigr({\bf R}^{3} 
- \left \{ 0 \right \} \Bigr)$
in the presence of a Coulomb potential provided that 
$|\gamma|$ (see (4.11)) is majorized by ${1\over 2}\sqrt{3}$,
as is shown in [16], following work by Weidmann
(see page 130 in [16] and references therein). In our paper,
however, we have focused on second-order differential operators,
and the consideration of a central potential, 
with the associated Hilbert space
$$
L^{2}({\bf R}_{+},r^{2}dr) \otimes L^{2}(S^{2},d\Omega)
$$
has eventually led to the second-order operators occurring in
(2.10) and (2.17) and acting on square-integrable functions on
the positive half-line. Our calculations, summarized in the points
(i)--(iv) above, remain therefore original. We should notice that 
the condition $|\gamma| < {1\over 2}\sqrt{3}$ 
found in [16] is compatible with
our inequalities (4.14) and (4.15) for all $l \geq 2$. In other words,
the condition on $\gamma$ ensuring essential self-adjointness of the 
Dirac operator leads also to essential self-adjointness of the 
second-order operators studied in our paper,
whereas the converse does not hold (one may find
a $|\gamma|$ smaller than $\sqrt{3}$ but greater than
${1\over 2}\sqrt{3}$). Our analysis
has possibly the merit of having shown that some extra care is
necessary when $l=0,1$, but this should not be unexpected, if one
bears in mind from section 3 that already in the free case the
values $l=0,1$ make it necessary to perform a separate analysis
(cf [20]).

We should also acknowledge that
in [21] the essential self-adjointness of powers of the
Dirac operator had been proved, but in cases when the potential 
$V$ is smooth. In particular, when the potential is a $C^{\infty}$ 
function on ${\bf R}^{3}$, no growth conditions on it are 
necessary to ensure essential self-adjointness of any power of
the Dirac operator [9, 21]. In our problem, however, we have
considered a Coulomb term in the potential, which is singular at
the origin. Although a regular solution of the eigenvalue problem
exists [8, 18], since the origin 
remains a Fuchsian singular point, the
domain of essential self-adjointness of the second-order operators
in the interacting case is changed. This is reflected by the
inequality (4.13) for the fulfillment of the limit-point condition
at zero, which now involves $\gamma$, and hence the atomic 
number [11, 12]. Note also that, to find a real-valued solution
which is regular at the origin in a Coulomb field, 
one only needs the weaker condition
$k^{2} \geq \gamma^{2}$ [11, 12]. 
Thus, a careful investigation of the
essential self-adjointness issue picks out a subset of the general
set of real-valued regular solutions.

For simplicity, we have considered in the end of section 4 only
one `linear' term. More precisely, however, two linear terms are
often studied, of scalar and vector nature, 
respectively [8]. Moreover, a naturally
occurring question is whether one can extend our 
qualitative analysis to study the (essential) self-adjointness
issue for operators involving the square root of the Laplacian
[22], i.e. $\sqrt{-c^{2}{\hbar}^{2} \bigtriangleup
+m_{0}^{2}c^{4}}-{Ze^{2}\over r}$.
Such problems have been the object of intensive 
investigations, but more work could be done from the point of
view of rigorous mathematical foundations.
In the light of the above remarks, there is some encouraging
evidence that new insight into the choice of phenomenological
central potentials can be gained by applying some powerful analytic
techniques along the lines described in our paper. In the near 
future, one might therefore hope to re-interpret from a deeper
perspective the previous work in the literature, including the
class of potentials responsible for quark confinement.
\vskip 0.3cm
\leftline {\bf Acknowledgments}
\vskip 0.3cm
\noindent
The authors are grateful to Romeo Brunetti for having brought
reference [16] to their attention.
This work has been partially supported by PRIN97 `Sintesi'.
\vskip 0.3cm
\leftline {\bf References}
\vskip 0.3cm
\noindent
\item {[1]}
Dirac P A M 1928 {\it Proc. R. Soc. Lond.} A {\bf 117} 610
\item {[2]}
Darwin C G 1928 {\it Proc. R. Soc. Lond.} A {\bf 118} 654;
{\it ibid.} A {\bf 120} 621
\item {[3]}
Gordon W 1928 {\it Z. Phys.} {\bf 48} 11
\item {[4]}
Kaburagi M, Kawaguchi M, Morii T, Kitazoe T and Morishita J 
1980 {\it Phys. Lett.} {\bf 97B} 143
\item {[5]}
Eichten E and Feinberg F 1981 {\it Phys. Rev.} D {\bf 23} 2724
\item {[6]}
Kaburagi M, Kawaguchi M and Morii T  
1981 {\it Z. Phys.} C {\bf 9} 213
\item {[7]}
Baker M, Ball J S and Zachariasen F 1995 {\it Phys. Rev.} D
{\bf 51} 1968
\item {[8]}
Franklin J 1998 `A simple Dirac wave function for a Coulomb 
potential with linear confinement' (HEP-PH 9812464)
\item {[9]}
Reed M and Simon B 1975 {\it Methods of Modern Mathematical 
Physics. II. Fourier Analysis and Self-Adjointness}
(New York: Academic)
\item {[10]}
Kato T 1995 {\it Perturbation Theory for Linear Operators}
(Berlin: Springer-Verlag)
\item {[11]}
Landau L D and Lifshitz E M 1971 {\it Relativistic Quantum 
Theory} (Oxford: Pergamon Press)
\item {[12]}
Greiner W 1990 {\it Relativistic Quantum Mechanics: Wave
Equations} (Berlin: Springer-Verlag)
\item {[13]}
Griesemer M and Lutgen J 1999 {\it J. Funct. Anal.} 
{\bf 162} 120
\item {[14]}
Whittaker E T and Watson G N 1927 {\it Modern Analysis}
(London: Cambridge University Press)
\item {[15]}
Reed M and Simon B 1980 {\it Methods of Modern Mathematical
Physics. I. Functional Analysis (Second Edition)} 
(New York: Academic)
\item {[16]}
Thaller B 1992 {\it The Dirac Equation} (Berlin: Springer--Verlag)
\item {[17]}
Eichten E and Gottfried K 1977 {\it Phys. Lett.} 
{\bf 66B} 286
\item {[18]}
Mur V D, Popov V S, Simonov A Yu and Yurov V P 1994
{\it JETP} {\bf 78} 1
\item {[19]}
Becirevic D and Le Yaouanc A 1999 {\it J. High Energy Phys.}
03(1999)021
\item {[20]}
Fewster C J 1993 `On the energy levels of the hydrogen atom'
(HEP-TH 9305102)  
\item {[21]}
Chernoff P R 1973 {\it J. Func. Anal.} {\bf 12} 401
\item {[22]}
Herbst I 1977 {\it Commun. Math. Phys.} {\bf 53} 285

\bye